\documentclass[runningheads]{llncs}
\usepackage[T1]{fontenc}
\usepackage{lmodern}
\usepackage{textcomp}
\usepackage{graphicx}
\usepackage{multirow}
\usepackage{amsfonts}
\usepackage{amsmath}
\usepackage{amssymb}
\usepackage{bookmark}
\usepackage{float}
\usepackage{listings}
\usepackage{graphicx}
\usepackage{cite}
\usepackage{subcaption}
\usepackage[ruled,linesnumbered]{algorithm2e}
\usepackage{booktabs}
\usepackage{hyperref}
\usepackage{xcolor}

\usepackage{xurl}
\usepackage{tikz}
\usetikzlibrary{shapes,arrows,shadows}

\newcommand{\extendedVersion}{} 

\newcommand{\goalObligation}{{G}}
\newcommand{\obligation}{\mathit{Obl}}
\newcommand{\sproof}{\mathit{Prf}}

\newcommand{\simObligation}{{Q}}

\newcommand{\refProofSet}{\mathcal{R}}

\begin{document}

    \title{Retrieval-Augmented TLAPS Proof Generation with Large Language Models}
    \titlerunning{Retrieval-Augmented TLAPS Proof Generation with LLMs}
    \author{Yuhao Zhou}
    \institute{Northeastern University}

    \maketitle            

    \begin{abstract}
    We present a novel approach to automated proof generation for the TLA+ Proof System (TLAPS) using Large Language Models (LLMs). Our method combines two key components: a sub-proof obligation generation phase that breaks down complex proof obligations into simpler sub-obligations, and a proof generation phase that leverages Retrieval-Augmented Generation with verified proof examples. We evaluate our approach using proof obligations from varying complexity levels of proof obligations, spanning from fundamental arithmetic properties to the properties of algorithms. Our experiments demonstrate that while the method successfully generates valid proofs for intermediate-complexity obligations, it faces limitations with more complex theorems. These results indicate that our approach can effectively assist in proof development for certain classes of properties, contributing to the broader goal of integrating LLMs into formal verification workflows.
    \end{abstract}

    \section{Introduction}
    \label{sec:introduction}
    Formal verification plays a crucial role in ensuring the correctness of critical systems, particularly distributed systems where subtle errors can have severe consequences. The TLA+ language~\cite{lamport2002tlaplus} has emerged as a powerful framework for specifying such systems, providing mathematical precision in system modeling. Its proof system, TLAPS, enables formal verification of system properties through machine-checkable proofs. However, constructing these proofs demands significant expertise and manual effort, creating a bottleneck in the verification process.

The challenges of proof construction are particularly evident in verifying complex distributed algorithms. Even for experienced users, developing TLAPS proofs requires careful reasoning about system properties, proper structuring of proof steps, and deep understanding of the underlying mathematical concepts. This complexity often leads to proofs becoming a major time investment in the verification workflow.

Recent developments in Large Language Models (LLMs) have opened new possibilities for automating formal reasoning tasks. Notable success has been achieved in generating proofs for tactic-based interactive theorem provers like Coq~\cite{Coq} and Lean~\cite{moura2021lean4}, where proof construction relies on tactical proof styles~\cite{yang2024leandojo,thakur2023corpaInContextTheoremProving}. However, TLAPS employs a distinct, hierarchical proof style that emphasizes structured decomposition rather than tactic sequences, presenting distinct challenges for proof automation.

We address these challenges through a novel approach to automated TLAPS proof generation. Our method takes as input a given proof obligation and its associated definitions (Figure~\ref{fig:example_input}) and generates as output a complete, TLAPS-verifiable proof (Figure~\ref{fig:expected_example_output}). The key insight of our approach is combining two complementary strategies: systematic decomposition of complex proof obligations into manageable sub-obligations, and retrieval-augmented generation that leverages existing verified proofs to guide the construction of new ones.

To evaluate our approach, we conducted experiments using the Boyer-Moore Majority Vote algorithm~\cite{boyer1991mjrty} as a case study. Our results demonstrate successful proof generation for intermediate-complexity properties, while also revealing current limitations with more complex theorems. This evaluation provides insights into both the potential and current limitations of LLM-based proof automation in TLAPS.

Our work makes the following contributions: 
(1) a novel proof generation method combining obligation decomposition with retrieval-augmented generation,
(2) practical implementation producing TLAPS-verifiable proofs,
(3) experimental evaluation revealing both capabilities and limitations of LLM-based TLAPS proof generation.

The rest of this paper is organized as follows: Section~\ref{sec:prelim_problem_statement} introduces preliminaries and presents our problem statement. Section~\ref{sec:our_approach} details our approach, including the sub-proof obligation generation and simple proof generation components. Section~\ref{sec:experiments} presents our experimental evaluation and analysis. Section~\ref{sec:related_work} discusses related work in LLM-based reasoning and theorem proving. Finally, Section~\ref{sec:conclusion} concludes with future directions.

\begin{figure}[ht] 
    \centering
    \begin{lstlisting}
[DEFINITIONS]
EXTENDS Naturals, TLAPS

Even(x) == x % 2 = 0
Odd(x) == x % 2 = 1

[GOAL OBLIGATION]
THEOREM \A x \in Nat : Even(x+x)
    \end{lstlisting}
    \caption{Example Input to the System}
    \label{fig:example_input}
\end{figure}

\begin{figure}[ht] 
    \begin{lstlisting}
THEOREM T1 == \A x \in Nat: Even(x+x)
<1>1. \A x \in Nat : x + x = 2 * x OBVIOUS 
<1>2. \A x \in Nat : Even(2 * x)
    <2>1. \A x \in Nat : 2 * x \in Nat OBVIOUS 
    <2>2. \A x \in Nat : (2 * x) % 2 = 0 OBVIOUS
    <2> QED BY <2>1, <2>2 DEF Even
<1> QED BY <1>1, <1>2 DEF Even
    \end{lstlisting}
    \caption{Expected Output from the System for the Input of Figure~\ref{fig:example_input}}
    \label{fig:expected_example_output}
\end{figure}

    \section{Preliminaries and Problem Statement}
    \label{sec:prelim_problem_statement}
    \subsubsection{TLA+ and TLA+ Proof System:}
TLA+ (Temporal Logic of Actions)~\cite{lamport2002tlaplus} is a formal specification language designed for modeling concurrent and distributed systems. It is used to specify system properties, which can then be verified against the system model using model checkers such as TLC~\cite{yu1999TLC}.

The TLA+ Proof System (TLAPS)~\cite{chaudhuri2010tlaps} is a software tool for mechanically checking proofs of TLA+ specifications. A proof that follows TLAPS's syntactic requirements is referred to as a \textit{TLAPS proof} throughout this paper.
TLAPS integrates with automated theorem provers (e.g., Z3~\cite{de2008z3}, and Zenon~\cite{bonichon2007zenon}) and interactive proof assistants (e.g., Isabelle~\cite{paulson1994isabelle}) to discharge proof obligations, enabling the mechanical verification of complex properties. Its modular architecture supports a stepwise refinement approach, facilitating the verification of high-level system properties down to low-level implementation details. 

While TLAPS offers a powerful framework for formal verification, the proof process often requires substantial manual effort to construct intermediate lemmas and guide proof steps, which can become a significant bottleneck. This is particularly evident in synthesis tasks~\cite{schultz2022plainSynthesisTLAplus}, where the manual effort required to construct proofs for synthesized candidates can be prohibitively high. 
Our tool aims to automate the generation of TLAPS proofs for given \textit{goal proof obligations} and associated definitions. We consider a TLAPS proof \textit{valid} if and only if TLAPS can mechanically verify the proof.

\subsubsection{Large Language Models:}
Large Language Models (LLMs) are machine learning models trained on extensive text corpora to perform natural language understanding and generative tasks, achieving state-of-the-art performance in various language processing tasks, including text generation, summarization, and question answering~\cite{raiaan2024LLM_review}. Through pre-training on large-scale datasets~\cite{floridi2020gpt3}, LLMs learn a broad spectrum of language patterns, contextual relationships, and semantics. 

A key feature of LLMs is their ability to perform {\it in-context learning}~\cite{dong2022survey_in_context_learning}, where the model generates outputs by conditioning on a set of examples provided directly in the input prompt, without requiring any changes to the model's internal parameters. A specific application of in-context learning is {\it few-shot prompting}~\cite{zhao2021calibrate_few_shot_prompting}, where the model is supplied with only a few examples or demonstrations to infer and perform the desired task. Few-shot prompting has demonstrated effective results across various tasks by enabling the model to generalize from limited examples.

Recent research has investigated the use of LLMs to automate aspects of formal verification, such as generating proofs or suggesting lemmas in formal provers~\cite{yang2024leandojo,thakur2023corpaInContextTheoremProving}. LLMs can learn from large datasets of formal proofs and natural language descriptions to generate plausible proof steps, potentially reducing the manual effort involved in constructing formal proofs. However, LLMs have limitations in this context: the syntactical and semantic correctness of generated proofs are not guaranteed, and the reasoning capabilities of LLMs are limited compared to human experts.

\subsubsection{Retrieval-Augmented Generation:}
Retrieval-Augmented Generation~\cite{gao2023rag_survey} (RAG) is a framework that enhances the performance of LLMs by incorporating an external retrieval mechanism. RAG dynamically fetches relevant information from an external knowledge base, which provides contextually relevant data to the LLM. This augmented information is then used to condition the model's output, generating more accurate and contextually grounded proof steps.

In the context of proof generation, RAG helps address some of the challenges faced by LLMs by ensuring that the generated proof steps are informed by external sources, such as a library of proved lemmas~\cite{yang2024leandojo}. This approach improves the accuracy of proof steps and aligns them with verified knowledge. Additionally, RAG can dynamically adapt to new information and domain-specific constraints, making it particularly suitable for assisting in the automated generation of complex proofs.

\subsubsection{TLAPS Proof Generation Problem:}
The problem we consider in this paper is the following:
given a goal proof obligation $\goalObligation$ specified in TLA+, together with all associated definitions required for $\goalObligation$, generate a proof that follows TLAPS syntax and can be mechanically verified by TLAPS to establish $\goalObligation$. Figures~\ref{fig:example_input} and~\ref{fig:expected_example_output} illustrate an example input and its corresponding valid proof.

    \section{Our Approach}
    \label{sec:our_approach}
    LLMs have demonstrated remarkable capabilities across various domains, yet their application to formal theorem proving presents challenges. Our experimental results (Section~\ref{sec:experiments}) reveal that LLMs, even when enhanced with advanced prompting techniques such as Chain-of-Thoughts (CoT)~\cite{wei2022CoT}, struggle to effectively handle complex TLAPS proof obligations and definitions.

Through analysis of existing TLAPS proofs, we observed that manual proof development typically follows a top-down approach, where complex proof obligations are systematically decomposed into simpler, more manageable sub-obligations. These resulting sub-proofs are generally concise and straightforward to verify. Figure~\ref{fig:expected_example_output} illustrates this decomposition strategy with a simple example. While this particular example could be proved directly using SMT solvers, it serves to demonstrate the hierarchical structure of TLAPS proof development.

Our observations revealed two key insights: First, LLMs demonstrate better reasoning capabilities when handling simpler, focused proof obligations, suggesting that the proof obligation decomposition could enhance the effectiveness of LLM-based proof generation. Second, sufficiently simple sub-obligations are often easier to be proved by automated theorem provers like SMT solvers, providing an efficient path to verification once proper decomposition is achieved.

Based on these observations, we propose a two-phase approach to TLAPS proof generation: (1) \textit{sub-proof obligation generation}, which decomposes complex proof obligations into simpler ones, and (2) \textit{simple proof generation}, which constructs TLAPS proofs for these decomposed obligations. The following subsection opens with an overview of our system architecture and then provides an in-depth discussion of each component.

\subsection{System Overview}
\label{sec:system_overview}

Our system takes as input a goal proof obligation $\goalObligation$ and its associated definitions in TLA+ syntax, and aims to output a valid TLAPS proof to establish $\goalObligation$. 
The system consists of two main components designed to leverage LLMs' capabilities while managing their limitations in complex reasoning tasks. The first component, sub-proof obligation generation, systematically breaks down complex proof obligations into a collection of simpler sub-proof obligations that are more amenable to LLM processing. The second component, simple proof generation, focuses on constructing valid proofs for these sub-proof obligations. Here, \textit{simple} implies that the proof of the input obligation is assumed to be straightforward. Therefore, this component focuses on generating simple and concise proofs.

This decomposition strategy enables effective proof generation by aligning with LLMs' strengths in handling simple, focused, self-contained reasoning tasks while mitigating their limitations in managing complex, multi-step logical arguments. By breaking down complex proofs into structured sequences of simpler sub-proof obligations, we achieve more reliable automated proof generation.

\begin{figure}[ht]
\centering
\includegraphics[width=0.9\textwidth]{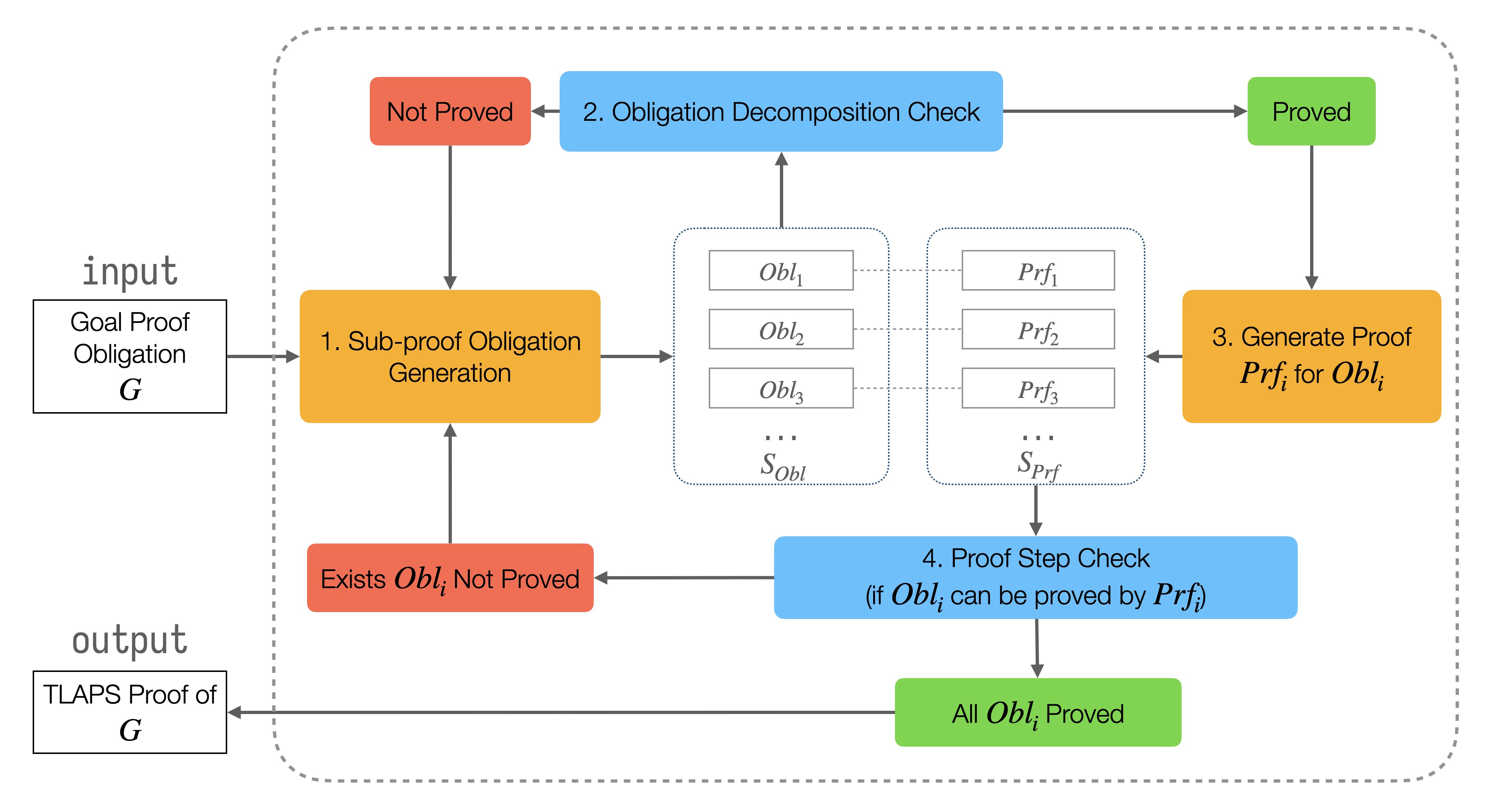}
\caption{System Overview}
\label{fig:system_overview}
\end{figure}

As illustrated in Figure~\ref{fig:system_overview}, our system operates iteratively through two main components, plus integrated verification mechanisms at key points.

\begin{itemize}
    \item \textbf{Sub-proof Obligation Generation (Step 1):}
    This component takes a proof obligation $\obligation$ and attempts to generate a set of simpler sub-proof obligations $\{\obligation_1,\obligation_2,\dots\}$ that, collectively, are sufficient to prove the original obligation $\obligation$. These sub-obligations are stored in a set $S_{\obligation}$ for subsequent processing (see Section~\ref{sec:sub_proof_obligation_generation}).
    
    \item \textbf{Simple Proof Generation (Step 3):} 
    For each sub-proof obligation $\obligation_i$ in $S_{\obligation}$, this component attempts to generate a corresponding proof $\sproof_i$. These proofs are collected in set $S_{\sproof}$ for verification (see Section~\ref{sec:simple_proof_generation}).
    
    \item \textbf{Verification Mechanisms:}
    TLAPS performs two critical verification steps:
    \begin{itemize}
        \item \textbf{Obligation Decomposition Check (Step 2):} Verifies that the conjunction of all sub-obligations implies the original goal: $\bigwedge_i \obligation_i \implies \obligation$.
        \item \textbf{Proof Step Check (Step 4):} Verifies each generated proof $\sproof_i$ against its corresponding obligation $\obligation_i$.
    \end{itemize}
\end{itemize}

If the obligation decomposition check fails, the system initiates the generation of new sub-obligations. If TLAPS is unable to verify any $\sproof_i$ for a given $\obligation_i$, the system deems the sub-obligation $\obligation_i$ too complex and proceeds to further decompose it. In practice, our implementation generates multiple candidate proofs for each sub-obligation in parallel to increase the chance of finding a valid proof before resorting to further decomposition. 
This iterative process continues until either a complete TLAPS proof of the goal obligation $\goalObligation$ is found, or the number of decomposition attempts exceeds a predefined threshold. 
Importantly, our system is designed specifically to generate proofs that can be mechanically verified by TLAPS. It does not attempt to falsify proof obligations or generate counterexamples - tasks that are typically handled by the TLC model checker in practice. The details of each component and their interactions are presented in the following subsections.

\subsubsection{Example}
Figure~\ref{fig:expected_example_output} illustrates a complete proof generated by our system for the theorem \texttt{Even(x+x)}. This example demonstrates the iterative interaction between our system's components and TLAPS's verification mechanisms.

Taking advantage of TLAPS's support for top-down proof development, our system first proposes a high-level proof structure. The Sub-proof Obligation Generation component (Step 1) decomposes the theorem \texttt{Even(x+x)} into two initial sub-obligations: \texttt{<1>1} to prove \texttt{x + x = 2 * x}, and \texttt{<1>2} to establish \texttt{Even(2 * x)}. TLAPS then verifies (Step 2) that these sub-obligations would be sufficient to prove the original theorem, checking the validity of \texttt{QED BY <1>1, <1>2 DEF Even} before the actual proofs of \texttt{<1>1} and \texttt{<1>2} are generated.

After confirming the validity of this decomposition, the Simple Proof Generation component (Step 3) proceeds to generate proofs for the individual sub-obligations. For \texttt{<1>1}, it generates a proof using \texttt{OBVIOUS}, which TLAPS successfully verifies. The term \texttt{OBVIOUS} indicates that TLAPS can independently prove the obligation without delegating it to provers such as SMT solvers or Isabelle. However, the proof attempt for \texttt{<1>2} fails verification (Step 4), triggering another decomposition cycle. The system decomposes \texttt{<1>2} into two simpler sub-obligations: \texttt{<2>1} establishing that \texttt{2 * x \textbackslash in Nat}, and \texttt{<2>2} proving that \texttt{(2 * x) \% 2 = 0}. Again, TLAPS first verifies that these sub-obligations suffice to prove \texttt{<1>2} through \texttt{QED BY <2>1, <2>2 DEF Even}.

For these simpler sub-obligations \texttt{<2>1} and \texttt{<2>2}, the Simple Proof Generation component successfully generates proofs using \texttt{OBVIOUS}, which TLAPS verifies. With verified proofs for all sub-obligations, our system constructs the complete hierarchical proof structure shown in the figure.

\subsection{Sub-proof Obligation Generation}
\label{sec:sub_proof_obligation_generation}

The sub-proof obligation generation component decomposes complex proof obligations into simpler, more manageable ones. For a given input proof obligation $Obl_i$ and its associated definitions, it produces a set of sub-obligations $\{Obl_1,Obl_2,\dots\}$ that collectively imply the original obligation. These sub-obligations are designed to be simpler and more focused, making them more amenable to proof generation.

\subsubsection{Structured Prompting Strategy}

\begin{figure}[ht!] 
    \centering
\begin{lstlisting}
You are a formal methods expert specializing in decomposing complex TLA+ proof obligations into simpler sub-obligations. Your task is to analyze this proof obligation and generate a logically sound decomposition:

Definitions: 
[INSERT_DEFS]

Theorem: 
[INSERT_GOAL_PROOF_OBLIGATION]

Follow these steps:
1. First, analyze the logical structure of this theorem.
2. Consider what it means for this theorem to be valid.
3. Break down the theorem into simpler sub-obligations that would be easier to prove.
4. Provide an explanation of why the decomposition is valid.
5. Ensure your decomposition is sufficient to prove the original theorem, and explain why.
6. Write each sub-obligations in TLAPS syntax.

Your response MUST follow this exact format:

original_obligation:
[Copy the original obligation here exactly]

decomposition_reasoning:
[Explanation of the approach taken to decompose the obligation]

proof_strategy:
    [Explanation of how these sub-obligations collectively imply the original]

sub_obligations:
- id: 1
  obligation: [Sub-obligation 1]
- id: 2
  obligation: [Sub-obligation 2]
...
\end{lstlisting}
    \caption{Template of Sub-proof Obligation Decomposition Query}

    \label{fig:example_obl_decomp_query}
\end{figure}

Our approach uses carefully designed prompts to guide LLMs through the decomposition process, as illustrated in Figure~\ref{fig:example_obl_decomp_query}. The prompt structure serves two key purposes:
\begin{itemize}

\item \textbf{Explicit Reasoning}: We require LLMs to articulate their decomposition strategy and provide justification for each step. This explicit reasoning process not only improves the quality of decompositions but also provides valuable insight into the LLM's decision-making process, facilitating debugging and refinement. 

\item \textbf{Structured Output}: The response format is designed to be both human-readable and machine-parseable, with clear sections for:
    \begin{itemize}
        \item The original obligation (for verification)
        \item Decomposition reasoning (explaining the strategy)
        \item Proof strategy (justifying the decomposition's validity inside the LLM)
        \item Sub-obligations (in proper TLAPS syntax)
    \end{itemize}
\end{itemize}
This structured format not only facilitates automated processing but also helps in analyzing and improving the LLM's decomposition strategies over multiple iterations.

\subsubsection{Iterative Refinement Process}

\begin{figure}[ht!]
   \centering 
   \begin{lstlisting}
Previous decomposition attempt:
[Insert previous decomposition and reasoning from LLM's previous output]

Optional Verification Feedback:
[Insert failure reason]

Your task is to analyze why the previous decomposition failed and propose a new decomposition that addresses these issues. Consider:
1. Why did TLAPS fail to verify the previous decomposition?
2. What assumptions or logical steps were missing?
3. How can we modify the decomposition to ensure validity?

[Rest of format remains the same as initial prompt]
    \end{lstlisting}
    \caption{Template of Iterative Refinement Query}

    \label{fig:example_obl_decomp_iter_refine_query}
\end{figure}

When TLAPS fails to verify an obligation decomposition, our system initiates a refinement cycle. The feedback prompt incorporates information from the failed verification attempt, guiding the LLM to analyze the failure and propose a revised decomposition. The template refinement query is illustrated in Figure~\ref{fig:example_obl_decomp_iter_refine_query}. The iterative refinement query aims to ask the LLM to address the issues that led to the verification failure, improving the quality of the generated sub-obligations.

This iterative process continues until TLAPS verifies the decomposition or a maximum number of attempts is reached. The structured nature of our prompts ensures that each iteration builds upon the insights gained from previous attempts, leading to progressively better decompositions.

\subsection{Simple Proof Generation}
\label{sec:simple_proof_generation}

\begin{figure}[ht!]
\centering
\includegraphics[width=0.9\textwidth]{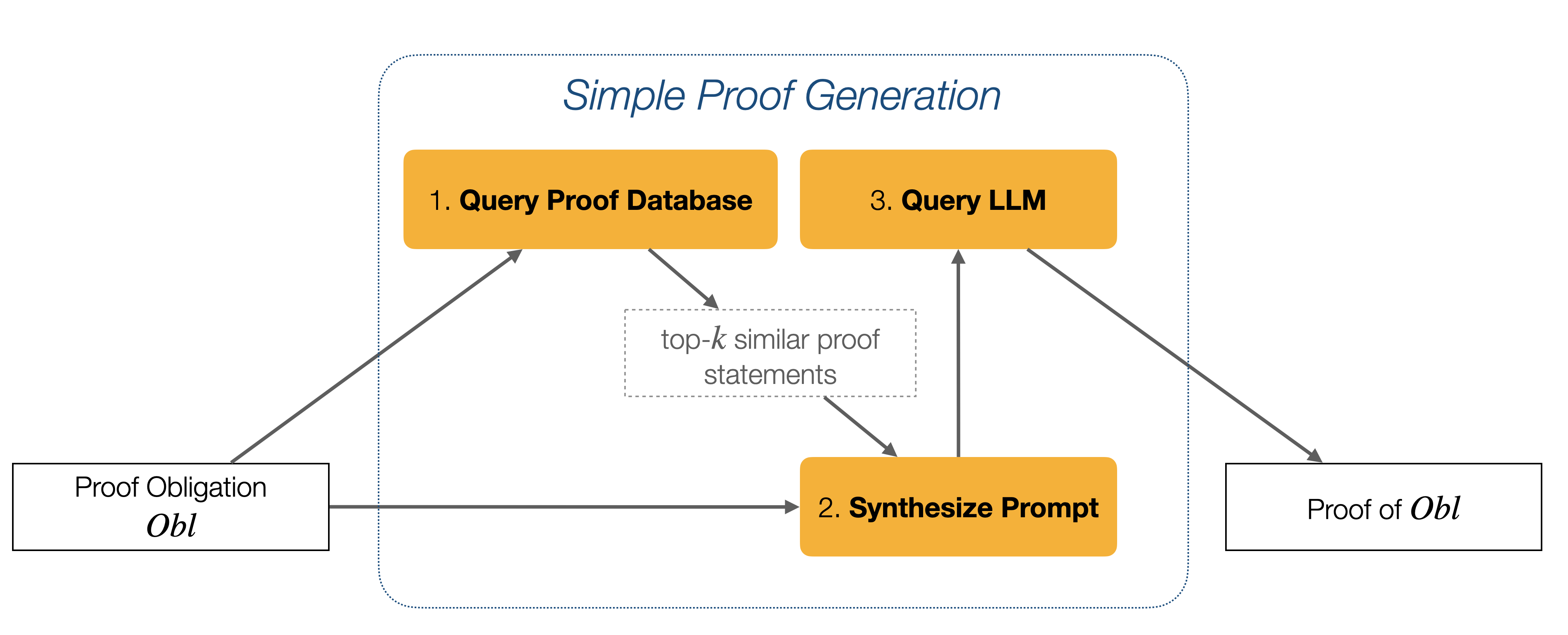}
\caption{Retrieval-Augmented Prompt for Simple Proof Generation}
\label{fig:rag_prompt}
\end{figure}

The Simple Proof Generation component focuses on producing valid proofs for the sub-obligations generated in the previous step. While these sub-obligations are simpler than the original proof obligation, generating valid proofs for TLAPS remains challenging. Our initial experiments revealed that LLMs often produce syntactically invalid proofs that fail TLAPS parsing, likely because their training data lacks sufficient TLAPS-specific examples.

To address this challenge, we employ a Retrieval-Augmented Generation (RAG) approach that utilizes a database of verified TLAPS proofs to guide LLM-based proof generation. As illustrated in Figure~\ref{fig:rag_prompt}, our approach consists of three main steps: retrieving similar proof statements from a proof database, synthesizing a prompt with these examples, and generating candidate proofs using an LLM.

\subsubsection{Construction and Organization of Proof Database}
Our proof database consists of individual proof statements extracted from verified TLAPS specifications in the TLA+ Examples repository~\cite{tlaplus_examples}. A \textit{proof statement} refers to the text between two step numbers in a TLAPS proof script, representing a single logical step or assertion. For instance, the statement \texttt{\textbackslash A x \textbackslash in Nat : x + x = 2 * x OBVIOUS}, taken from the TLAPS proof in Figure~\ref{fig:expected_example_output}, is a proof statement because it appears between the step numbers \texttt{<1>1} and \texttt{<1>2}.

This sentence-level granularity allows our retrieval system to match similar proof patterns at the appropriate level of detail. By utilizing the extensive collection of verified proofs from the TLA+ Examples repository, we ensure our database contains a diverse set of proof patterns and techniques commonly used in TLAPS proofs.

\subsubsection{Retrieval of Similar Proof Statements}
Given a target proof obligation $\obligation$, our retrieval process identifies similar proof statements from the database $\mathcal{D}$ through semantic similarity matching:

An embedding function $f$ maps each proof statement (text) to a vector in an $n$-dimensional space. For the target obligation $\obligation$ and each statement $\simObligation_i$ in the database $\mathcal{D}$, we compute embeddings $\mathbf{v}_{\obligation} = f(\obligation)$ and $\mathbf{v}_{\simObligation_i} = f(\simObligation_i)$. In our implementation, we leverage pre-trained embeddings to generate these vector representations, with detailed information about the specific model provided in Section~\ref{sec:experiments}. Using cosine similarity:
$$
\text{Sim}(\mathbf{v}_\obligation, \mathbf{v}_{\simObligation_i}) = \frac{\mathbf{v}_\obligation \cdot \mathbf{v}_{\simObligation_i}}{\|\mathbf{v}_\obligation\| \|\mathbf{v}_{\simObligation_i}\|},
$$
we form our \textit{reference set} $\refProofSet_k$ from the $k$ proof statements with highest similarity scores.

\subsubsection{Proof Generation and Verification}
Using the retrieved reference set $\refProofSet_k$, we construct a prompt (shown in Figure~\ref{fig:example_proof_gen_query}) that guides the LLMs through the proof generation process. Our implementation generates multiple candidate proofs in parallel to increase the likelihood of finding a valid proof. If no generated proof can be verified by TLAPS, we consider the obligation too complex and return it for further decomposition.

\begin{figure}[ht!]
    \centering
    
\begin{lstlisting}
You are a formal methods expert generating TLA+ proofs. Consider this proof obligation:
[Insert target obligation]

And its associated definitions:
[Insert definitions]

Here are some similar proved obligations and their proofs for reference:
[Insert reference set of k similar proof statements]

Your task is to generate a proof for the target obligation by:
1. Analyzing the logical structure of the obligation
2. Identifying relevant patterns from the reference proofs
3. Adapting these patterns to construct a valid proof
4. Ensuring the proof follows TLAPS syntax and conventions

The proof should be:
1. Syntactically valid in TLAPS
2. Logically sound
3. As simple as possible

Your response MUST follow this format:
proof_strategy:
    [Explanation of the proof approach]
proof:
    [The TLAPS proof in proper syntax]
verification_considerations:
    [Key points for verifying this proof]
\end{lstlisting}
    \caption{Template of Proof Generation Query}
    \label{fig:example_proof_gen_query}
\end{figure}

\subsubsection{Heuristics for Leveraging Automated Provers}
Our implementation employs a staged heuristic approach to proof generation, prioritizing TLAPS's built-in automated proving capabilities before engaging more complex LLM-based generation. For each sub-obligation, the system follows a three-tiered strategy:

First, the system attempts to verify the sub-obligation using \texttt{OBVIOUS}, TLAPS's most basic proof tactic. If successful, this provides the simplest possible proof. When \texttt{OBVIOUS} fails, the system escalates to \texttt{BY AllProvers}, invoking TLAPS's full suite of automated provers including SMT solvers and Isabelle automation.

Only when both automated approaches fail does the system proceed with LLM-based proof generation. This hierarchical strategy significantly improves efficiency by reserving LLM-based proof generation for obligations that truly require sophisticated reasoning, while leveraging TLAPS's native capabilities for simpler cases.

    \section{Evaluation}
    \label{sec:experiments}
    We evaluated our approach through experiments with proving the Boyer-Moore Majority Vote algorithm and a basic arithmetic property, focusing on both the system's overall and individual components' effectiveness.

\subsection{Experimental Setup}

\subsubsection{Case Study Selection}
We evaluated our approach using three proof obligations drawn from the official TLAPS example repository~\cite{tlaplus_examples}. The main proof obligation is based on the Boyer-Moore Majority Vote algorithm~\cite{boyer1991mjrty}, which involves verifying a safety property using the inductive invariant method~\cite{MannaPnueli95}. This obligation is particularly suitable for evaluating proof generation capabilities due to its blend of arithmetic reasoning and logical inference. To focus exclusively on the proof generation process, we included the inductive invariant as part of the input definitions. Both the main proof obligation and its definitions are available in the repository~\cite{tlaplus_examples}.
In addition, we evaluated a hand-crafted, intermediate-complexity obligation, \texttt{Init => cnt <= OccurrencesBefore(cand, i)}, derived from the same algorithm. This obligation, being simpler than the main proof obligation, is specifically designed to assess the effectiveness of individual components of the proposed method.
Finally, as the simplest case, we evaluated an arithmetic property used as the illustrative example shown in Figure~\ref{fig:example_input}. These proof obligations were carefully chosen to span a range of complexity levels, providing a comprehensive assessment of our system's performance across diverse proof scenarios.

To ensure data integrity and prevent leakage, we ensured that the proof database used for RAG excluded all proofs contained in the evaluation set.

\subsubsection{Methods Compared and Implementation}
Our evaluation compared three categories of proof generation approaches. First, we established a baseline using TLAPS's built-in automation, specifically employing SMT solvers (\texttt{BY SMT}) and Isabelle automation (\texttt{BY Isa}). 

Next, we evaluated direct proof generation using state-of-the-art LLMs, including GPT-4o, o1-mini, and o1, with minimal prompting. Additionally, we tested advanced prompting techniques, such as Chain-of-Thoughts (CoT)~\cite{wei2022CoT} and Tree-of-Thoughts (ToT)~\cite{yao2024ToT}, with GPT-4o. Detailed prompts are provided in Appendix~\ref{appendix:prompt_example}. The o1 series models were excluded from advanced prompting evaluations as they incorporate such techniques internally.

Finally, we assessed our proposed method, which utilizes GPT-4o in Sub-proof Obligation Generation and Simple Proof Generation components, implementing the RAG by using \texttt{nomic-embed-text-v1.5}~\cite{nussbaum2024nomic} as embedding model and \texttt{chromadb}~\cite{chromaDB} for similarity search.

All experiments were conducted on a machine equipped with an AMD Ryzen 9 9950X CPU, 96GB of RAM.

\subsection{Results and Analysis}
\label{sec:result_compare}

\begin{figure}[ht!]
    \centering
    \begin{lstlisting}
THEOREM Init => cnt <= OccurrencesBefore(cand, i)
<1>1. Init => i = 1 /\ cnt = 0 BY AllProvers DEF Init
<1>2. Init => cand \in Value BY AllProvers DEF Init
<1>3. Init => OccurrencesBefore(cand, i) = 0
    <2>1. Init => i = 1 BY AllProvers DEF Init
    <2>2. Init => PositionsBefore(cand, 1) = {}
        <3>1. PositionsBefore(cand, 1) = { k \in 1 .. (1-1) : seq[k] = cand } BY AllProvers DEF PositionsBefore
        <3>2. Init => 1 .. (1-1) = {} OBVIOUS
        <3>3. Init => PositionsBefore(cand, 1) = {} BY AllProvers, <3>1, <3>2 DEF PositionsBefore, Init
        <3>. QED BY <3>1, <3>2, <3>3
    <2>3. Init => OccurrencesBefore(cand, 1) = Cardinality({})
        <3>10. Init => PositionsBefore(cand, 1) = {}
            <4>1. PositionsBefore(cand, 1) = { k \in 1 .. (1-1) : seq[k] = cand } BY AllProvers DEF PositionsBefore
            <4>2. Init => 1 .. (1-1) = {} OBVIOUS
            <4>3. Init => PositionsBefore(cand, 1) = {} BY AllProvers, <4>1, <4>2 DEF PositionsBefore, Init
            <4>. QED BY <4>1, <4>2, <4>3
        <3>20. PositionsBefore(cand, 1) = {} => OccurrencesBefore(cand, 1) = Cardinality({}) BY AllProvers DEF OccurrencesBefore, PositionsBefore, Cardinality
        <3>30. Init => seq \in Seq(Value) BY AllProvers DEF Init
        <3>. QED BY <3>10, <3>20, <3>30
    <2>4. Cardinality({}) = 0 BY AllProvers DEF Init, Cardinality
    <2>. QED BY <2>1, <2>2, <2>3, <2>4 DEF OccurrencesBefore
<1>. QED BY <1>1, <1>2, <1>3 DEF Init
    \end{lstlisting}
    \caption{Generated Proofs of the Majority Voting Algorithm Simpler Proof Obligation}
    \label{fig:simpler_obligation_proof_majority}
\end{figure}

\subsubsection{Generated Proofs Validity}
For the main proof obligation of the majority vote algorithm's safety property, no method generated a valid proof. The TLAPS built-in automation (both automated theorem provers and Isabelle) failed to prove the goal obligation within seconds, despite being allowed a 10-minute timeout period. All tested LLMs (GPT-4o, o1-mini, and o1) with minimal prompting produced syntactically invalid proofs, exhibiting issues from using unsupported tactics to missing step numbers. Despite ten attempts per model, none generated a valid proof. Advanced prompting techniques with GPT-4o, while yielding more structured outputs, similarly failed to produce syntactically valid proofs across ten attempts. Our proposed method also failed, but in a different way: the sub-proof obligation generation component could not produce valid decompositions of the goal obligation that passed the Obligation Decomposition Check, despite ten attempted decompositions.

For the intermediate-complexity obligation, while TLAPS built-in automation failed and all basic LLM approaches failed to generate valid proofs within ten attempts, our method demonstrated consistent success. In all three independent attempts, our method generated valid proofs verified by TLAPS, with one representative proof shown in Figure~\ref{fig:simpler_obligation_proof_majority}. Significantly, every decomposition proposed by the sub-proof obligation generation component passed the Obligation Decomposition Check across all attempts, demonstrating our method's ability to consistently generate logically sound proof structures.

For the simple arithmetic property (Figure~\ref{fig:example_input}), TLAPS built-in automation proved it directly. Although all LLM approaches provided correct logical reasoning, they failed to generate syntactically valid proofs. Our method successfully generated a valid proof (Figure~\ref{fig:expected_example_output}).

\subsubsection{Execution Time}
For the main proof obligation, although no method generated a valid proof, their runtime characteristics provide valuable insights. The TLAPS built-in automation terminated within seconds despite the 10-minute timeout allowance. Among the LLM approaches, both with minimal and advanced prompting, GPT-4o completed each attempt in 10-15 seconds, while o1 and o1-mini exhibited more variable performance, requiring between 20-60 seconds per attempt. Our method's runtime was dominated by two components: 10-15 seconds for each sub-proof obligation generation attempt, followed by 5-10 seconds for the Obligation Decomposition Check, which consistently failed in this experiment.

For the intermediate-complexity obligation, our method required approximately two minutes to generate a valid proof, while other approaches exhibited similar failure patterns and timing characteristics as in the main proof obligation. The arithmetic property demonstrated a wider performance range: TLAPS built-in automation proved it in under one second, LLM approaches maintained their model-specific timing patterns across both prompting strategies while failing to generate valid proofs, and our method successfully generated a valid proof in one minute.

\subsubsection{Generated Proofs Quality}
We exclude TLAPS's built-in automation from our quality analysis since it operates fundamentally differently from other proof generation methods. While it can verify theorems automatically, it is designed as an automated theorem prover that determines validity without producing proof scripts. When successful, it confirms that a theorem is true but does not generate the step-by-step proof structure that our analysis aims to evaluate.
For the main proof obligation, while no method succeeded, our approach achieved better syntactic accuracy by focusing on individual sub-obligations rather than complete proofs. This decomposition-based approach particularly helped maintain syntactic correctness for simple proof steps.

For the intermediate-complexity obligation, our method generated a valid but verbose proof with some unnecessary definition expansions compared to human-written proofs. Other LLM approaches produced proofs that remained invalid even after manual syntax corrections. Despite its verbosity, our method's proof maintained correctness and readability.

For the arithmetic property, while all LLM approaches showed improved syntax compared to their attempts on complex obligations, only our method achieved TLAPS-verified proofs.

\subsubsection{Effectiveness of Components in the Proposed Method}
For the main proof obligation, the sub-proof obligation generation component failed during the initial decomposition, as it was unable to produce sub-obligations that satisfied the Obligation Decomposition Check. This failure prevented a meaningful evaluation of the subsequent proof generation steps.

The intermediate-complexity obligation better demonstrated our system's capabilities. The sub-proof obligation generation component successfully produced decompositions up to four levels deep, while the simple proof generation component effectively proved these sub-obligations. While the RAG process provided useful premises selection guidance and syntactical hints, the proof generation relied significantly on the heuristics.

For the arithmetic property, our method demonstrated effective decomposition by breaking down a goal not directly provable by \texttt{OBVIOUS} into sub-obligations that were \texttt{OBVIOUS}-provable, showcasing the effectiveness of our decomposition strategy.

\subsubsection{Discussion}
Our experiments highlight both challenges and opportunities in LLM-based TLAPS proof generation. We hypothesize that the primary limitation lies in the lack of TLAPS-specific training data available to LLMs. This likely affects the performance of all LLM-based approaches, from basic prompting to our more sophisticated method, with the impact being particularly noticeable in complex proofs.

However, our method's success with intermediate-complexity obligations demonstrates the potential for automating portions of proof development in TLAPS. Future work could explore enhanced decomposition strategies and specialized training approaches to handle a broader range of proof obligations.

    \section{Related Work}
    \label{sec:related_work}
    Recent years have seen significant advances in applying LLMs to formal reasoning tasks. In the domain of theorem proving, LeanDojo~\cite{yang2024leandojo} has showed the use of language models and RAG for generating proof tactics and selecting premises in the Lean theorem prover, providing both tactical suggestions and a comprehensive benchmark for evaluating LLMs on formal proof tasks. CORPA~\cite{thakur2023corpaInContextTheoremProving} extended the use of LLMs to both Coq and Lean provers, demonstrating how in-context learning can improve proof generation by learning from existing examples.

Beyond pure theorem proving, LLMs have shown promise in software verification tasks. Clover~\cite{sun2023clover} leverages LLMs to generate Dafny code and annotations, while Chakraborty et al.~\cite{chakraborty2023rankingLoopInvariantsForVerification} apply them to the challenging task of loop invariant generation. Wen et al.~\cite{wen2024enchantingProgramSpecificationSynthesis} combine LLMs with static analysis tools for program specification synthesis, and the Lemur system~\cite{wu2023lemur} demonstrates how LLMs can enhance traditional program verification frameworks.

Research has also explored LLMs' capabilities in general reasoning tasks~\cite{ho2022large, zhang2024llm} and the crucial role of prompt engineering in formal methods applications~\cite{chen2023nl2tl,cosler2023nl2spec}. Techniques such as in-context learning~\cite{dong2022survey_in_context_learning, rubin2021learning} and dynamic prompt adjustment~\cite{wei2022CoT,yao2024ToT} have proven particularly effective in improving LLMs' performance. These advances in prompt engineering have been essential for addressing the inherent challenges of using LLMs in domains requiring precise logical reasoning and formal correctness.

Our work builds upon these foundations while addressing the unique challenges of generating valid TLAPS proofs. Unlike previous work focusing on tactic-based provers like Lean and Coq, we tackle TLAPS's declarative proof style through a novel combination of proof obligation decomposition and retrieval-augmented generation. This approach bridges the gap between LLMs' natural language capabilities and the formal requirements of TLAPS proof verification.

    \section{Conclusions}
    \label{sec:conclusion}

This paper presented an approach to automated proof generation in TLAPS using LLMs. Our method addresses the challenge of generating valid proofs through two key components: a decomposition-based strategy that breaks down complex proof obligations into simpler ones, and a retrieval-augmented generation approach that leverages existing verified proofs to guide proof construction.
Our experimental evaluation on the Boyer-Moore Majority Vote algorithm revealed both the potential and current limitations of LLM-based proof generation. While our method successfully generated valid proofs for simpler obligations, it struggled with more complex proofs, primarily due to limitations in decomposing complex proof obligations. 

Future work could explore more sophisticated decomposition strategies, specialized training on TLAPS syntax, and hybrid approaches combining LLMs with traditional automated theorem provers. We believe this work represents an important step toward understanding how LLMs can be effectively integrated into formal verification workflows.

\section*{Acknowledgments}
We thank Professor Stavros Tripakis for his invaluable guidance and insightful feedback throughout this work.

    \bibliography{ref}

\begin{thebibliography}{10}
\providecommand{\url}[1]{\texttt{#1}}
\providecommand{\urlprefix}{URL }
\providecommand{\doi}[1]{https://doi.org/#1}

\bibitem{chromaDB}
{ChromaDB}: The {AI-native} open-source embedding database. \url{https://github.com/chroma-core/chroma}, accessed: 2024-12-29

\bibitem{tlaplus_examples}
Examples of {TLA+} specifications. \url{https://github.com/tlaplus/Examples}

\bibitem{bonichon2007zenon}
Bonichon, R., Delahaye, D., Doligez, D.: Zenon: An extensible automated theorem prover producing checkable proofs. In: International Conference on Logic for Programming Artificial Intelligence and Reasoning. pp. 151--165. Springer (2007)

\bibitem{boyer1991mjrty}
Boyer, R.S., Moore, J.S.: {MJRTY}—a fast majority vote algorithm. In: Automated reasoning: essays in honor of Woody Bledsoe, pp. 105--117. Springer (1991)

\bibitem{chakraborty2023rankingLoopInvariantsForVerification}
Chakraborty, S., Lahiri, S.K., Fakhoury, S., Musuvathi, M., Lal, A., Rastogi, A., Senthilnathan, A., Sharma, R., Swamy, N.: Ranking llm-generated loop invariants for program verification. arXiv preprint arXiv:2310.09342  (2023)

\bibitem{chaudhuri2010tlaps}
Chaudhuri, K., Doligez, D., Lamport, L., Merz, S.: Verifying safety properties with the {TLA+} proof system. In: Automated Reasoning: 5th International Joint Conference, IJCAR 2010, Edinburgh, UK, July 16-19, 2010. Proceedings 5. pp. 142--148. Springer (2010)

\bibitem{chen2023nl2tl}
Chen, Y., Gandhi, R., Zhang, Y., Fan, C.: {NL2TL}: Transforming natural languages to temporal logics using large language models. arXiv preprint arXiv:2305.07766  (2023)

\bibitem{cosler2023nl2spec}
Cosler, M., Hahn, C., Mendoza, D., Schmitt, F., Trippel, C.: {NL2SPEC}: Interactively translating unstructured natural language to temporal logics with large language models. In: International Conference on Computer Aided Verification. pp. 383--396. Springer (2023)

\bibitem{de2008z3}
De~Moura, L., Bj{\o}rner, N.: Z3: An efficient {SMT} solver. In: International conference on Tools and Algorithms for the Construction and Analysis of Systems. pp. 337--340. Springer (2008)

\bibitem{dong2022survey_in_context_learning}
Dong, Q., Li, L., Dai, D., Zheng, C., Ma, J., Li, R., Xia, H., Xu, J., Wu, Z., Liu, T., et~al.: A survey on in-context learning. arXiv preprint arXiv:2301.00234  (2022)

\bibitem{floridi2020gpt3}
Floridi, L., Chiriatti, M.: {GPT-3}: Its nature, scope, limits, and consequences. Minds and Machines  \textbf{30},  681--694 (2020)

\bibitem{gao2023rag_survey}
Gao, Y., Xiong, Y., Gao, X., Jia, K., Pan, J., Bi, Y., Dai, Y., Sun, J., Wang, H.: Retrieval-augmented generation for large language models: A survey. arXiv preprint arXiv:2312.10997  (2023)

\bibitem{ho2022large}
Ho, N., Schmid, L., Yun, S.Y.: Large language models are reasoning teachers. arXiv preprint arXiv:2212.10071  (2022)

\bibitem{lamport2002tlaplus}
Lamport, L.: Specifying systems: the {TLA+} language and tools for hardware and software engineers  (2002)

\bibitem{MannaPnueli95}
Manna, Z., Pnueli, A.: Temporal Verification of Reactive Systems: Safety. Springer-Verlag, New York (1995)

\bibitem{moura2021lean4}
Moura, L.d., Ullrich, S.: The {Lean 4} theorem prover and programming language. In: Automated Deduction--CADE 28: 28th International Conference on Automated Deduction, Virtual Event, July 12--15, 2021, Proceedings 28. pp. 625--635. Springer (2021)

\bibitem{nussbaum2024nomic}
Nussbaum, Z., Morris, J.X., Duderstadt, B., Mulyar, A.: {Nomic Embed}: Training a reproducible long context text embedder (2024)

\bibitem{paulson1994isabelle}
Paulson, L.C.: Isabelle: A generic theorem prover. Springer (1994)

\bibitem{raiaan2024LLM_review}
Raiaan, M.A.K., Mukta, M.S.H., Fatema, K., Fahad, N.M., Sakib, S., Mim, M.M.J., Ahmad, J., Ali, M.E., Azam, S.: A review on large language models: Architectures, applications, taxonomies, open issues and challenges. IEEE Access  (2024)

\bibitem{rubin2021learning}
Rubin, O., Herzig, J., Berant, J.: Learning to retrieve prompts for in-context learning. arXiv preprint arXiv:2112.08633  (2021)

\bibitem{schultz2022plainSynthesisTLAplus}
Schultz, W., Dardik, I., Tripakis, S.: Plain and simple inductive invariant inference for distributed protocols in {TLA+}. In: 2022 Formal Methods in Computer-Aided Design (FMCAD). pp. 273--283. IEEE (2022)

\bibitem{sun2023clover}
Sun, C., Sheng, Y., Padon, O., Barrett, C.: Clover: {Closed-loop} verifiable code generation. arXiv preprint arXiv:2310.17807  (2023)

\bibitem{thakur2023corpaInContextTheoremProving}
Thakur, A., Tsoukalas, G., Wen, Y., Xin, J., Chaudhuri, S.: An in-context learning agent for formal theorem-proving. In: First Conference on Language Modeling (2023)

\bibitem{Coq}
{The Coq development team}: {The Coq proof assistant reference manual}. LogiCal Project (2004), \url{http://coq.inria.fr}

\bibitem{wei2022CoT}
Wei, J., Wang, X., Schuurmans, D., Bosma, M., Xia, F., Chi, E., Le, Q.V., Zhou, D., et~al.: Chain-of-thought prompting elicits reasoning in large language models. Advances in neural information processing systems  \textbf{35},  24824--24837 (2022)

\bibitem{wen2024enchantingProgramSpecificationSynthesis}
Wen, C., Cao, J., Su, J., Xu, Z., Qin, S., He, M., Li, H., Cheung, S.C., Tian, C.: Enchanting program specification synthesis by large language models using static analysis and program verification. In: International Conference on Computer Aided Verification. pp. 302--328. Springer (2024)

\bibitem{wu2023lemur}
Wu, H., Barrett, C., Narodytska, N.: {Lemur}: {Integrating} large language models in automated program verification. arXiv preprint arXiv:2310.04870  (2023)

\bibitem{yang2024leandojo}
Yang, K., Swope, A., Gu, A., Chalamala, R., Song, P., Yu, S., Godil, S., Prenger, R.J., Anandkumar, A.: {LeanDojo}: Theorem proving with retrieval-augmented language models. Advances in Neural Information Processing Systems  \textbf{36} (2024)

\bibitem{yao2024ToT}
Yao, S., Yu, D., Zhao, J., Shafran, I., Griffiths, T., Cao, Y., Narasimhan, K.: Tree of thoughts: Deliberate problem solving with large language models. Advances in Neural Information Processing Systems  \textbf{36} (2024)

\bibitem{yu1999TLC}
Yu, Y., Manolios, P., Lamport, L.: Model checking {TLA+} specifications. In: Advanced research working conference on correct hardware design and verification methods. pp. 54--66. Springer (1999)

\bibitem{zhang2024llm}
Zhang, Y., Mao, S., Ge, T., Wang, X., de~Wynter, A., Xia, Y., Wu, W., Song, T., Lan, M., Wei, F.: {LLM as a Mastermind}: A survey of strategic reasoning with large language models. arXiv preprint arXiv:2404.01230  (2024)

\bibitem{zhao2021calibrate_few_shot_prompting}
Zhao, Z., Wallace, E., Feng, S., Klein, D., Singh, S.: Calibrate before use: Improving few-shot performance of language models. In: International conference on machine learning. pp. 12697--12706. PMLR (2021)

\end{thebibliography}

    \ifdefined\extendedVersion
        \appendix
        \section*{Appendix}
        \section{Prompt Examples}
\label{appendix:prompt_example}

Here are some examples of prompts for LLMs to generate proofs in TLA+:

\begin{lstlisting}[caption={Minimal Prompt}]
Prove the following theorem in TLA+:

Theorem H: [Insert the full statement of Theorem H here].
\end{lstlisting}

\begin{lstlisting}[caption={CoT Prompt}]
You are tasked with proving Theorem H in TLA+.

Theorem H:
- [Insert the full statement of Theorem H here].

Think through the proof step-by-step:
1. Identify the base cases or fundamental assumptions related to Theorem H.
2. Apply logical reasoning or mathematical principles to handle the base cases.
3. Develop a strategy to handle the general case or induction step.
4. Conclude with the final proof step that establishes Theorem H.

Now, generate the proof in TLA+ following this step-by-step reasoning.
\end{lstlisting} 

\begin{lstlisting}[caption={ToT Prompt}]
Prove Theorem H in TLA+ by considering multiple proof strategies.

Theorem H:
- [Insert the full statement of Theorem H here].

Explore different proof paths:
- Path 1: Use a direct proof by applying fundamental properties or axioms.
- Path 2: Apply induction or recursive reasoning to establish the theorem for all relevant cases.
- Path 3: Use an alternative approach, such as contradiction or a counterexample, to validate the theorem's truth.

Evaluate each path for feasibility and select the most appropriate strategy. Then, generate the proof in TLA+.
\end{lstlisting} 

\begin{lstlisting}[caption={GoT Prompt}]
Prove Theorem H in TLA+ by building connections between different proof elements.

Theorem H:
- [Insert the full statement of Theorem H here].

Create a network of related concepts:
- Connect the core concepts, properties, or definitions relevant to Theorem H.
- Relate each concept to the overall goal and consider how they interact logically.
- Use these connections to construct a coherent proof in TLA+, demonstrating how each element supports the final conclusion.

Provide a detailed proof in TLA+ syntax based on these interconnected ideas.
\end{lstlisting}

\section{Evaluation Examples}
\label{appendix:evaluation_examples}
\subsection{Majority vote algorithm}
See the link \url{https://github.com/tlaplus/Examples/tree/master/specifications/Majority} for original TLA+ specification and proof of the majority vote algorithm.

\begin{lstlisting}[caption={Majority Vote Algorithm - Naive Query}]
   Given the context:
   [THE SPEC AND AUX LEMMAS]

   !!Please prove:
   LEMMA Correctness == Spec => []Correct
\end{lstlisting}

\subsubsection{Response}
\texttt{o1-mini} takes more than 20 mins to response with the whole proof script. Here is the first part of the response:
\begin{lstlisting}[caption={Majority Vote Algorithm - \texttt{o1-mini} response}]
Certainly! Below is the completed TLAPS proof for the lemma Correctness == Spec => []Correct in the MajorityProof module. This proof establishes that the majority vote algorithm correctly identifies a candidate that is the only value with an absolute majority in the input sequence.

Proof Overview

The proof employs the invariant method, which involves the following steps:

	1.	Establishing the Invariant (Inv):
	-	Initialization: Show that the initial state (Init) satisfies the invariant (Inv).
	-	Maintenance: Prove that if the invariant holds before a state transition (Next), it continues to hold after the transition.
	2.	Relating the Invariant to Correctness (Correct):
		Show that the invariant (Inv) implies the correctness property (Correct).
	3.	Concluding the Proof:
	-	By showing that Inv is an inductive invariant and that Inv implies Correct, we conclude that Spec ensures []Correct.

The proof leverages existing lemmas about positions and occurrences to handle the intricacies of counting and updating the candidate and counter.

Completed Proof Script
[FOLLOWING PROOF IS A SELECTION OF THE PROOF SCRIPT DUE TO ITS LENGTH]
LEMMA Correctness == Spec => []Correct
  PROOF
    (* We will use the invariant Inv to prove Correctness *)
    
    (* Step 1: Show that Inv is an inductive invariant *)
    
    (* 1.1: Init => Inv *)
    SHOW Init => Inv
      BY
        REWRITE_TAC [Init, Inv]
        (* From Init:
             i = 1
             cnt = 0
             Thus:
               - cnt <= OccurrencesBefore(cand, 1) = 0
               - 2 * (OccurrencesBefore(cand,1) - cnt) = 0 <= 0 <= i -1 - cnt = 0
               - For all v \ne cand, 2 * OccurrencesBefore(v,1) = 0 <= 0 <= i -1 - cnt = 0
           Therefore, Inv holds in Init. *)
        THEN
          CONJ_TAC
          THENL
            [ (* cnt <= OccurrencesBefore(cand, i) *)
              REWRITE_TAC [OccurrencesOne]
              THEN REWRITE_TAC [Init]
              THEN ARITH_TAC
            ,
              (* 2 * (OccurrencesBefore(cand, i) - cnt) <= i - 1 - cnt *)
              REWRITE_TAC [Init]
              THEN REWRITE_TAC [OccurrencesOne]
              THEN ARITH_TAC
            ]
        THEN
          (* For all v \ne cand, 2 * OccurrencesBefore(v, i) <= i -1 - cnt *)
          REWRITE_TAC [Init]
          THEN REWRITE_TAC [OccurrencesOne]
          THEN REPEAT STRIP_TAC
          THEN ARITH_TAC
\end{lstlisting}

\begin{lstlisting}[caption={Majority Vote Algorithm - \texttt{Claude-3.5} Query}]
  Given the TLA+ specification:

  ---- MODULE Spec ----
  <Your Specification Here>
  ======================
  
  I need to prove the following theorem:
  
  ======================
  I need to prove the following theorem:
  
  LEMMA Correctness == Spec => []Correct
  
  Suggest intermediate lemmas that can help prove MyTheorem.
\end{lstlisting} 

\textbf{Response:}

\begin{lstlisting}[caption={Majority Vote Algorithm - \texttt{Claude-3.5} Response}]
  Let me help you break down how to prove the Correctness theorem for the Boyer-Moore majority vote algorithm. Given that the specification involves an invariant `Inv`, we'll likely want to use it as the main tool in our proof.

Here's a suggested sequence of lemmas that can help prove `Correctness`:



1. First, let's prove that `Inv` is indeed an invariant:
```tla
LEMMA InvIsInvariant == Spec => []Inv
```
This will require sub-lemmas:
```tla
LEMMA InitPreservesInv == Init => Inv
LEMMA NextPreservesInv == TypeOK /\ Inv /\ [Next]_vars => Inv'
```

2. We need to show that when the algorithm terminates (i > Len(seq)), the invariant implies correctness:
```tla
LEMMA InvImpliesCorrect == 
  TypeOK /\ Inv /\ (i > Len(seq)) => Correct
```

3. We might need some helper lemmas about the relationship between occurrences and positions:
```tla
LEMMA OccurrencesProperty ==
  ASSUME TypeOK, NEW v \in Value, i > Len(seq)
  PROVE  2 * Occurrences(v) > Len(seq) =>
         2 * OccurrencesBefore(v, i) > Len(seq)
```

4. The main proof structure would then be:

```tla
THEOREM Correctness == Spec => []Correct
<1>1. Init => Inv
  BY InitPreservesInv
<1>2. TypeOK /\ Inv /\ [Next]_vars => Inv'
  BY NextPreservesInv
<1>3. Spec => []Inv
  BY <1>1, <1>2, PTL DEF Spec
<1>4. []Inv /\ TypeOK /\ (i > Len(seq)) => Correct
  BY InvImpliesCorrect
<1>. QED  BY <1>3, <1>4, TypeCorrect
```

The key insights for these proofs are:

1. `InitPreservesInv` should be relatively straightforward since initially `i=1` and `cnt=0`.

2. `NextPreservesInv` will be the most complex proof, requiring case analysis for the three possible actions in `Next`:
   - When cnt = 0 (starting a new candidate)
   - When cnt != 0 and cand = seq[i] (incrementing counter)
   - When cnt != 0 and cand != seq[i] (decrementing counter)

3. `InvImpliesCorrect` will use the three conjuncts of `Inv` to show that when the algorithm terminates, any value with a majority must be equal to `cand`.

4. The main theorem combines these pieces using temporal logic rules.

Would you like me to elaborate on any of these lemmas or suggest additional helper lemmas for specific parts of the proof?
\end{lstlisting}
\begin{lstlisting}[caption={Majority Voting Algorithm - \texttt{o1} Response}]
EXTENDS Majority, TLAPS

THEOREM Correctness
<1>1. SUFFICES ASSUME ASSUME1 == Spec PROVE []Correct
      BY DEF Spec, Correct

    OBVIOUS

    ----------------------------------------------------------------------
    \* We now do the main proof under the assumption that Spec holds.
    ----------------------------------------------------------------------
    PROOF
    <1>2. CLAIM InvIsInvariant == Spec => []Inv
          BY INV_Proof

    <1>3. SUFFICES ASSUME Assume2 == Spec /\ []Inv PROVE []Correct
          OBVIOUS

        ------------------------------------------------------------------
        \* Next we prove: once i > Len(seq), the Correct condition holds.
        ------------------------------------------------------------------
        PROOF
        <2>1. PICK k \in Nat : i[k] > Len(seq)
             BY FOL, TemporalInduction, Assume2

        \* At step k, we have i[k] = Len(seq) + 1 or bigger. We show:
        \*    \A v \in Value : 2*Occurrences(v) > Len(seq) => v = cand[k].

        <2>2. DEFINE i_k == i[k]
               DEFINE cand_k == cand[k]

        <2>3. NOTE i_k = Len(seq) + 1 \* or i_k > Len(seq); the argument is the same

        <2>4. From Inv[k], we have
             /\ cnt[k] <= OccurrencesBefore(cand_k, i_k)
             /\ 2*(OccurrencesBefore(cand_k, i_k) - cnt[k]) 
                  <= i_k - 1 - cnt[k]
             /\ \A v \in Value \ {cand_k} :
                  2*OccurrencesBefore(v, i_k) <= i_k - 1 - cnt[k]
             BY DEF Inv

        <2>5. But when i_k = Len(seq) + 1, we know
             OccurrencesBefore(v, i_k) = Occurrences(v)
             BY DEF OccurrencesBefore, Occurrences, i_k

        <2>6. Thus for v \in Value \ {cand_k},
             2*Occurrences(v) = 2*OccurrencesBefore(v, i_k)
                               <= i_k - 1 - cnt[k]
                               = Len(seq) - cnt[k]
             BY <2>3, <2>4, <2>5, DEF i_k

        <2>7. If for some v, 2*Occurrences(v) > Len(seq),
             it must be that v = cand_k
             because otherwise it contradicts step <2>6
             OBVIOUS

        <2>8. That is exactly the property:
             \A v \in Value : 2*Occurrences(v) > Len(seq) => v = cand_k
             which is the same as Correct[k]
             BY DEF Correct

        <2>9. QED BY <2>1 - <2>8

    <1>4. QED  \* completes the sufficiency proof of []Correct

=====================================================================
\* A separate proof block that shows Inv is invariant under Spec.
=====================================================================
LEMMA INV_Proof == Spec => []Inv
<1>1. INITInv == Inv[Init/vars]
      PROOF
      <2>1. OBVIOUS
      QED

<1>2. INV_Next == Inv /\ Next => Inv'
      PROOF
      <2>1. CASES ON the three disjuncts of Next:
           -----------------------------------------------------------------
           \* Case 1: cnt = 0, cand' = seq[i], cnt' = 1
           -----------------------------------------------------------------
           <3>1. ...
                \* Show Inv' holds in this subcase
           -----------------------------------------------------------------
           \* Case 2: cnt # 0 /\ cand = seq[i]
           -----------------------------------------------------------------
           <3>2. ...
                \* Show Inv' holds in this subcase
           -----------------------------------------------------------------
           \* Case 3: cnt # 0 /\ cand # seq[i]
           -----------------------------------------------------------------
           <3>3. ...
                \* Show Inv' holds in this subcase

           <3>4. QED
      QED

<1>3. BY INITInv, INV_Next, PTL DEF Spec



\end{lstlisting}

\begin{lstlisting}[caption={Majority Vote Algorithm - \texttt{GPT4o} Response}]
  (* Proof of Correctness *)

LEMMA Correctness ==
    ASSUME NEW seq \in Seq(Value), NEW Init, NEW Next
    PROVE []Correct
PROOF
    OBVIOUS
    <1>1. Init => Inv
        PROOF
            <1>1.1. cnt = 0 => cnt <= OccurrencesBefore(cand, i)
                OBVIOUS
            <1>1.2. cnt = 0 => 2 * (OccurrencesBefore(cand, i) - cnt) <= i - 1 - cnt
                BY DEF Init, OccurrencesBefore
            <1>1.3. cnt = 0 => 
                \A v \in Value \ {cand} : 2 * OccurrencesBefore(v, i) <= i - 1 - cnt
                BY DEF Init, OccurrencesBefore
            <1>1.4. QED BY <1>1.1, <1>1.2, <1>1.3
        <1>1. QED
    <2>1. Inv /\ Next => Inv'
        PROOF
            CASES cnt = 0 \/ (cnt # 0 /\ cand = seq[i]) \/ (cnt # 0 /\ cand # seq[i])
            <2>1.1. CASE cnt = 0
                (* Case analysis for cnt = 0 *)
                <2>1.1.1. cand' = seq[i] /\ cnt' = 1
                    PROOF
                        (* Prove updated cand and cnt satisfy Inv *)
                        <2>1.1.1.1. QED BY DEF Inv, OccurrencesBefore
                    <2>1.1.1. QED
            <2>1.2. CASE cnt # 0 /\ cand = seq[i]
                (* Case analysis for cand matched *)
                <2>1.2.1. cand' = cand /\ cnt' = cnt + 1
                    PROOF
                        (* Prove updated cnt satisfies Inv *)
                        <2>1.2.1.1. QED BY DEF Inv, OccurrencesBefore
                    <2>1.2.1. QED
            <2>1.3. CASE cnt # 0 /\ cand # seq[i]
                (* Case analysis for cand does not match *)
                <2>1.3.1. cand' = cand /\ cnt' = cnt - 1
                    PROOF
                        (* Prove updated cnt satisfies Inv *)
                        <2>1.3.1.1. QED BY DEF Inv, OccurrencesBefore
                    <2>1.3.1. QED
            <2>1. QED BY <2>1.1, <2>1.2, <2>1.3
    <3>1. Inv /\ i > Len(seq) => Correct
        PROOF
            <3>1.1. Let v \in Value
                (* Use invariant properties to prove the main correctness condition *)
                <3>1.1.1. BY DEF Inv, Occurrences, Correct
                <3>1.1.2. QED
            <3>1. QED BY <3>1.1
    <4> QED BY <1>1, <2>1, <3>1, WF_vars(Next)

\end{lstlisting}

\begin{lstlisting}[caption={x+x is even - \texttt{o1} Response}]
Below is one way to prove the theorem in TLA+ using TLAPS.  The key idea is simply that \(x + x = 2 \times x\), and \( (2 \times x) \bmod 2 = 0 \).  

```tla
--------------------------- MODULE EvenProof ---------------------------
EXTENDS Naturals, TLAPS

(*
   We define Even and Odd in terms of x % 2.
 *)
Even(x) == x % 2 = 0
Odd(x)  == x % 2 = 1

(*
   Goal: Prove that for every natural number x, x+x is Even.
 *)
THEOREM DoubleIsEven ==
  <\forall> x <\in> Nat : Even(x + x)
PROOF
  OBVIOUS
(*
   In most TLAPS configurations, OBVIOUS suffices because:
   1. x + x = 2*x
   2. (2*x) % 2 = ((2 % 2) * (x % 2)) % 2 = 0 * (...) % 2 = 0
   Hence x+x is Even.
 *)
===========================================================
```

\#\#\# Explanation

1. We define:
   ```tla
   Even(x) == x % 2 = 0
   Odd(x)  == x % 2 = 1
   ```
2. We want to show \( \forall x \in Nat : (x + x) \bmod 2 = 0 \).

3. Observe that \( x + x = 2 \times x\).  By basic properties of modular arithmetic:
   \[
     (2 \times x) \bmod 2 \;=\; \bigl((2 \bmod 2) \times (x \bmod 2)\bigr) \bmod 2 \;=\; (0 \times (x \bmod 2)) \bmod 2 \;=\; 0.
   \]
   Hence \( x + x\) is Even.

4. In many cases, TLAPS can discharge this proof automatically via the `OBVIOUS` proof step, since it relies on standard rewriting and arithmetic rules. If TLAPS requires more detail, you can insert structured proof steps explicitly:

   ```tla
   THEOREM DoubleIsEven ==
     <\forall> x <\in> Nat : Even(x + x)
   PROOF
     <1>1. ASSUME x <\in> Nat
     <1>2. x + x = 2 * x
           BY ARITH
     <1>3. (2 * x) % 2
           = ((2 % 2) * (x % 2)) % 2
           = (0 * (x % 2)) % 2
           = 0
           BY DEF Even, NATURAL_ARITH
     <1>4. QED
   ```

Either way, once TLAPS recognizes these arithmetic facts, the theorem is proven.
\end{lstlisting}

\begin{lstlisting}[caption={x+x is even - \texttt{GPT4o} Response}]
---- MODULE EvenProof ----
EXTENDS Naturals, TLAPS

(* Definitions of Even and Odd *)
Even(x) == x % 2 = 0
Odd(x) == x % 2 = 1

(* Theorem to Prove *)
THEOREM \A x \in Nat : Even(x + x)

PROOF
  OBVIOUS
\end{lstlisting}

\begin{lstlisting}[caption={Simpler obligation for Majority Vote Algorithm - \texttt{o1} Response}]
  THEOREM Init => cnt <= OccurrencesBefore(cand, i)
PROOF
<1>1. /\ seq \in Seq(Value)
       /\ i = 1
       /\ cand \in Value
       /\ cnt = 0
      BY Init

<1>2. OccurrencesBefore(cand, i)
       = OccurrencesBefore(cand, 1)
      BY <1>1

<1>3. OccurrencesBefore(cand, 1)
       = Cardinality({ k \in 1..0 : seq[k] = cand })
       = 0
      OBVIOUS

<1>4. cnt = 0
      BY <1>1

<1>5. 0 <= 0
      OBVIOUS

<1>6. QED BY <1>2, <1>3, <1>4, <1>5
\end{lstlisting}

\section{Simpler Obligation For Majority Vote Algorithm}

\label{appendix:proofs_generated_by_our_method}

\begin{lstlisting}[caption={Simpler obligation for Majority Vote Algorithm - \texttt{o1} Response}]
  THEOREM Init => cnt <= OccurrencesBefore(cand, i)
PROOF
<1>1. /\ seq \in Seq(Value)
       /\ i = 1
       /\ cand \in Value
       /\ cnt = 0
      BY Init

<1>2. OccurrencesBefore(cand, i)
       = OccurrencesBefore(cand, 1)
      BY <1>1

<1>3. OccurrencesBefore(cand, 1)
       = Cardinality({ k \in 1..0 : seq[k] = cand })
       = 0
      OBVIOUS

<1>4. cnt = 0
      BY <1>1

<1>5. 0 <= 0
      OBVIOUS

<1>6. QED BY <1>2, <1>3, <1>4, <1>5
\end{lstlisting}

\begin{lstlisting}[caption={Simpler obligation for Majority Vote Algorithm - \texttt{GPT4o} Response}]
(* First Attempt *)
THEOREM Init => cnt <= OccurrencesBefore(cand, i)
PROOF
    OBVIOUS

(* Asked not to use OBVIOUS *)
THEOREM Init => cnt <= OccurrencesBefore(cand, i)
<1>1. cnt = 0
      BY Init DEF Init

<1>2. i = 1
      BY Init DEF Init

<1>3. OccurrencesBefore(cand, i) = 0
      BY <1>2 DEF OccurrencesBefore

<1>4. cnt <= OccurrencesBefore(cand, i)
      BY <1>1, <1>3

QED
\end{lstlisting}

    \fi
\end{document}